\documentclass[a4paper,11pt]{article}
\pdfoutput=1 
\usepackage{jinstpub} 
\usepackage{multirow}
\usepackage[above, below]{placeins}
\usepackage{acronym}
\usepackage{amsmath}
\usepackage{hyperref}
\usepackage{graphicx}
\usepackage{lineno}
\usepackage{siunitx}
\usepackage{subcaption}
\usepackage{textgreek}

\title{Time-assisted energy reconstruction in a highly-granular hadronic calorimeter}

 \author[1]{C.~Graf\note{now at Robert Bosch GmbH, Renningen, Germany.}}
 \author[2]{and F.~Simon\note{Corresponding author.}}
 \affiliation{Max-Planck-Institut f\"ur Physik,\\F\"ohringer Ring 6, 80805 M\"unchen, Germany}

\emailAdd{fsimon@mpp.mpg.de}

\abstract{The software compensation algorithms developed for the CALICE Analog Hadron Calorimeter are extended to incorporate time information on the cell level, and the performance is studied in GEANT4 simulations with a detector model of a highly-granular SiPM-on-tile calorimeter. The addition of nanosecond-level time resolution is found to result in significant improvement of the energy resolution by approximately \SIrange{3}{4}{\percent} for local software compensation compared to the software compensation based on local energy density alone, with further improvement possible with better timing resolution. The high correlation of energy density and time variables show that both provide sensitivity to correlated underlying shower features, which limits the potential of timing information when used as a global rather than a local variable.}

\begin{document}
\maketitle
\flushbottom

%\linenumbers

\acrodef{ASIC}{application-specific integrated circuit}
\acrodef{AHCAL}{Analog Hadron Calorimeter}
\acrodef{EBU}{Ecal Base Unit}
\acrodef{FWHM}{full width at half maximum}
\acrodef{HBU}{Hcal Base Unit}
\acrodef{MC}{Monte Carlo}
\acrodef{MPPC}{Multi-Pixel Photon Counter}
\acrodef{RMS}{root mean squared}
\acrodef{SiPM}{Silicon Photomultiplier}
\acrodef{PMT}{Photomultiplier Tube}
\acrodef{LTP}{large technological prototype}
\acrodef{PFA}{particle-flow algorithms}

\acused{ASIC}
\acused{FWHM}
\acused{MPPC}
\acused{RMS}
%\acused{SiPM}
\acused{PMT}

%\linenumbers

\section{Introduction}
\label{sec:Intro}

Highly-granular electromagnetic and hadronic calorimeters are a crucial part of particle-flow-centred detectors at future $e^+e^-$ colliders \cite{Thomson:2009rp,Brient:2001fow,pfaMorgunov}. 
The \ac{AHCAL}
developed by the CALICE collaboration represents one concept for a highly-granular hadronic calorimeter \cite{CALICE:2010fpb}. It is a sampling calorimeter on plastic scintillator basis read out
by \acp{SiPM} with cell sizes of $\num{3} \times \SI{3}{\cm^2}$.

While the spatial resolution and the separation of showers
inside a particle jet play a key role in particle flow detectors \cite{Thomson:2009rp}, the energy
resolution of the hadronic calorimeter still adds to the overall detector performance, both via the improvement of the energy measurement of neutral hadrons, and through improved cluster-track matching in \ac{PFA} due to more accurate energy estimates of clusters assigned to charged particles \cite{Tran:2017tgr}.
In longitudinally and transversely segmented non-compensating hadronic calorimeters the energy resolution for hadronic showers can be improved by amplitude or energy-density dependent offline weighting of the measured signals in individual detector cells \cite{Abramowicz:1980iv,H1CalorimeterGroup:1993uzc,ATLASLiquidArgonEMECHEC:2004hkk}, exploiting the difference in response to electromagnetic and purely hadronic components of the showers. The high granularity of the CALICE calorimeters makes them perfectly suited for such reconstruction techniques, which have been developed under the term {\it software compensation} \cite{Adloff:2012gv} for highly-granular calorimeters. For the \ac{AHCAL}, an improvement of the hadronic energy resolution of approximately 20\% compared to a reconstruction without such techniques was achieved. 

Since the large-scale physics prototypes constructed by CALICE in the mid-2000s, the technology of highly-granular calorimeters has evolved, both on the sensor side with improved performance and better device-to-device uniformity, and in the area of front-end electronics, most notably by providing cell-by-cell time measurements. This enables the full five-dimensional reconstruction of particle showers in space, local energy density or amplitude, and time. The \ac{LTP} of the \ac{AHCAL} \cite{Sefkow:2018rhp}, a 22\,000 channel, 0.5 m$^3$ SiPM-on-tile highly-granular hadronic calorimeter with steel absorbers, which was first operated in beam tests with hadrons in 2018, has the ability to provide hit-time measurements on the cell level with nanosecond precision, as demonstrated in preliminary studies \cite{Emberger:2021lsz}. This capability may allow conclusions about the amount of slow neutrons generated in the shower, since measurements with different absorber materials have shown that delayed components in the signals in hadronic calorimeters are primarily due to the neutron component in the shower. Examples that show this connection of neutron activity and late signals are  measurements in  steel and uranium calorimeters \cite{WA78:1985num, Caldwell:1992te}, in a scintillating fiber calorimeter with lead absorber \cite{Acosta:1990bq} and in a copper-based dual readout calorimeter \cite{Akchurin:2007uf}. Shower energy that is absorbed in the generation of neutrons during the spallation processes to overcome the nuclear binding energy is lost for calorimetric purposes. Thus estimating the amount of neutrons in a given shower may result in an improvement of the energy measurement. The extension of the software compensation techniques used for highly-granular calorimeters to include time as an additional dimension is thus of high interest to explore the potential in this area. 

In the following, methods are presented for incorporating the timing capabilities of a highly-granular hadronic calorimeter into the energy reconstruction. For this, two different approaches to software compensation, a global and a local one, are presented, extending the techniques developed in \cite{Adloff:2012gv} to include the time domain in addition to purely energy-based observables. All methods are developed and evaluated on simulated data.

\section{Detector geometry and simulated data}

The detector geometry used for the simulation is a sampling calorimeter with steel
absorber and 60 active layers. The design 
follows that of the CALICE AHCAL \ac{LTP}, but with an increased number of layers to reduce the effect of longitudinal energy leakage on the energy reconstruction. The active elements of the calorimeter are  $\SI{3}{} \times \SI{3}{\cm^2}$ polystyrene-based scintillator tiles with the scintillation light read out by \acp{SiPM} housed on electronics boards with integrated front-end ASICs \cite{Sefkow:2018rhp}.
Each active layer consists of 576 channels, resulting in a transverse size of $\SI{72}{}\times\SI{72}{\cm^2}$. The total stainless steel absorber thickness per layer is \SI{17}{mm}, with \SI{16}{mm} provided by the mechanical absorber structure itself, and \SI{1}{mm} from the cassettes housing the active elements. The \ac{LTP} has 38 layers, which is not sufficiently deep to fully contain higher-energy hadronic showers without an additional tail catcher with coarser longitudinal sampling, or an electromagnetic calorimeter upstream of the hadronic calorimeter. To avoid energy-leakage-induced distortions of the energy resolution studies discussed in the present paper, 60 active layers and the corresponding absorbers have been simulated. The simulation
was performed using the DD4HEP \cite{Frank:2014zya} framework in combination with GEANT4 v10.03 \cite{Agostinelli:2002hh}. The \texttt{QGSP\_BERT\_HP} physics list was used for
all GEANT4 simulation samples discussed in this study. Scintillator saturation was implemented using a Birks' parameter of $k_B\,=\,0.07943\,\SI{}{mm MeV^{-1}}$ \cite{Hirschberg:1992xd}.

The idealized energy depositions generated by GEANT4 are combined per detector cell, assuming a maximum integration time of \SI{2000}{\ns}, and are transformed into hits in a digitization process that mimics the detector effects. %This includes effects of the read-out electronics (i.e., signal shaping and electronic noise) and of the \acp{SiPM} (i.e, statistical smearing and saturation). 
The influence of the \acp{SiPM} response on the signal is simulated via discretization into photoelectrons and Poissonian smearing, as well as a simulation of the saturation effects originating from the finite number of pixels in the sensor. The front-end electronics are taken into account via a simulation of shaping effects by combination of energy deposits within 50 ns into a single hit, and the addition of electronics noise, which manifests as additional smearing of the signal amplitude. The energy depositions in units of \SI{}{\keV} are converted to the MIP scale by a factor of \SI{477}{\keV/\text{MIP}}. This factor is calibrated such that a minimum ionizing particle (MIP) (i.e., a muon in the \SI{}{\GeV} range) has a most probable energy deposition of \SI{1}{\text{MIP}} in a scintillator tile traversed at normal incidence. Detector cells with an energy deposition below \SI{0.5}{\text{MIP}} are not considered in the analysis, in line with the threshold typically applied in the analysis of CALICE \ac{AHCAL} data. For the time measurements a gaussian smearing with a resolution of \SI{1}{\ns} is applied to the generator-level time stamp of the hit, which is given by the time of the first energy deposit in the cell. 

The CALICE collaboration has observed good agreement between test beam data of its
physics prototype and simulated hadronic showers with respect to the hit energy
distributions and shower shapes \cite{Bilki:2014bga}. It was shown by the CALICE T3B Experiment that the time distribution of hadronic showers, as simulated by GEANT4 with the physics list used in the present study, is in good agreement with the simulations using a steel absorber \cite{Adloff:2014rya}. 

\section{Correlation of energy and time observables}
\label{sec:correlations}

\begin{figure}[ht]
  \captionsetup[subfigure]{aboveskip=-2pt}
  \center
  \begin{subfigure}{0.515\textwidth}
    \includegraphics[width = 1\textwidth]{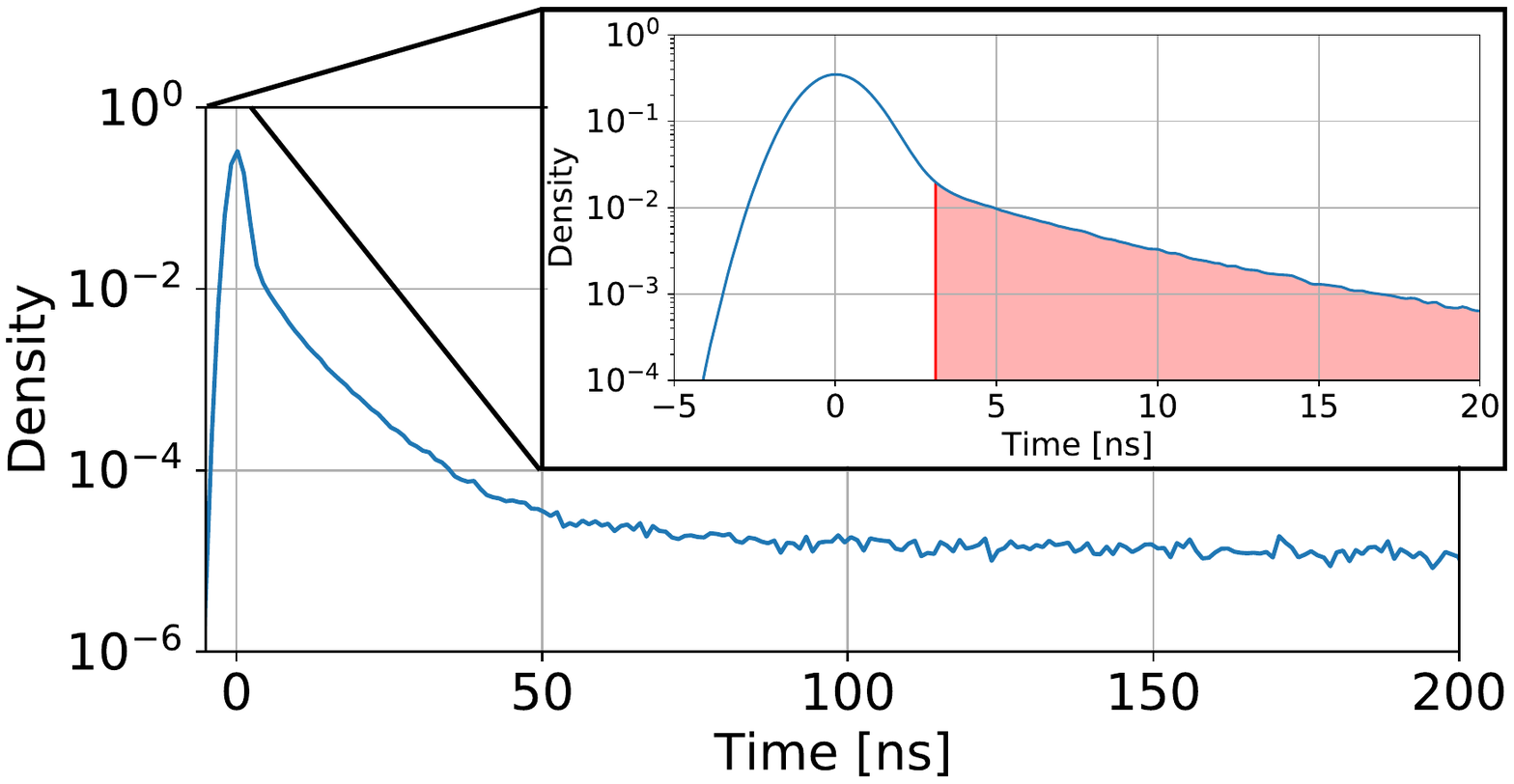}
    \vspace{-33pt}
    \caption{}
    \label{fig:TimeDistribution}
  \end{subfigure}
  \hfill
  \begin{subfigure}{0.475\textwidth}
    \includegraphics[width = 1\textwidth]{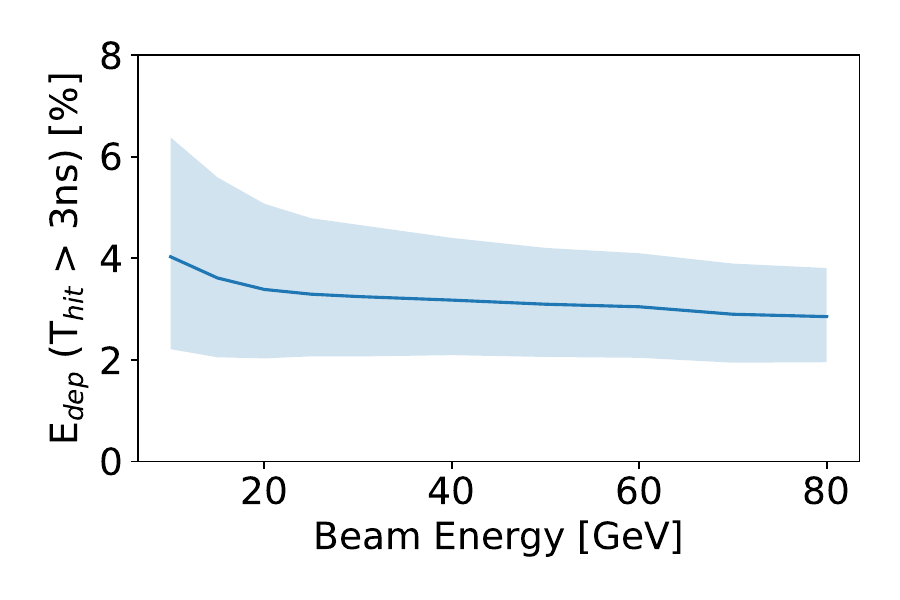}
    \caption{}
    \label{fig:mean_eDepLateHits1}
  \end{subfigure}
  \caption{\textbf{(a)} Hit time distribution for \SI{40}{\GeV} pions in the simulated
  calorimeter on a larger and smaller time scale. For the smaller time scale, the
  hit times above \SI{3}{\ns}, which are called \emph{late} throughout this study,
  are indicated in red. \textbf{(b)} Median of the fraction of measured late
  energy deposistions in a shower. The shaded region indicates the \SI{50}{\percent}
  quantile.}
  \label{fig:timeObservables}
\end{figure}

The overall hit time distribution for \SI{40}{\GeV} pions is shown in
\autoref{fig:TimeDistribution} on two different time scales. The zero time is calibrated such that a muon hitting a cell would produce a time stamp of $t=0$. Thus, the time of flight of the particles between the layers is already taken into account in the calibration. Time stamps below zero are due to the smearing introduced by the finite time resolution. In \autoref{fig:TimeDistribution} three different regimes can be identified: The instantaneous
part up to about \SI{3}{\ns}, governed by the time resolution of the detector, an intermediate, exponentially
falling, part mainly due to hits caused by neutron elastic scattering, and a late
tail dominated by neutron capture events. In the present study, the fraction of energy deposited later than \SI{3}{\ns} is adopted as the variable to characterize the timing behaviour of an event. The hits contributing to this fraction are indicated in red in \autoref{fig:TimeDistribution}.
In \autoref{fig:mean_eDepLateHits1} the median value of this fraction is shown
over the range of beam energies considered.
The shaded region visualizes the central \SI{50}{\percent}-interval. The median
value is about \SI{3}{\percent} for most beam energies and slightly higher for
lower energies. The monotonic decrease of this value is expected, as the electromagnetic
fraction is increasing for higher energies, resulting in a reduction of the influence of the neutron-dominated later parts of the shower.

\begin{figure}[ht]
\captionsetup[subfigure]{aboveskip=-7pt}
  \center
  \begin{subfigure}{0.495\textwidth}
    \includegraphics[width = 1\textwidth]{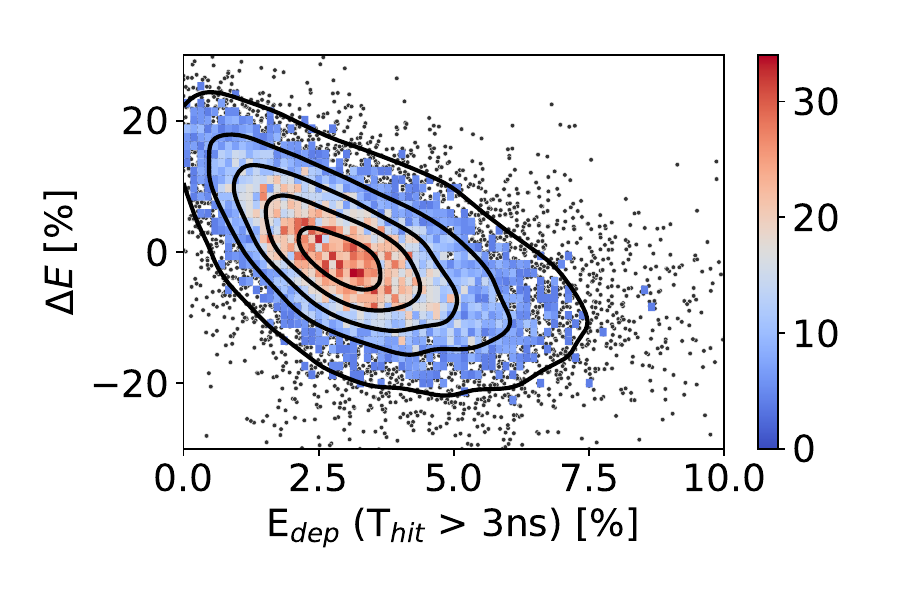}
    \caption{}
    \label{fig:Tedep_2dCorrelation}
  \end{subfigure}
  \hfill
  \begin{subfigure}{0.495\textwidth}
    \includegraphics[width = 1\textwidth]{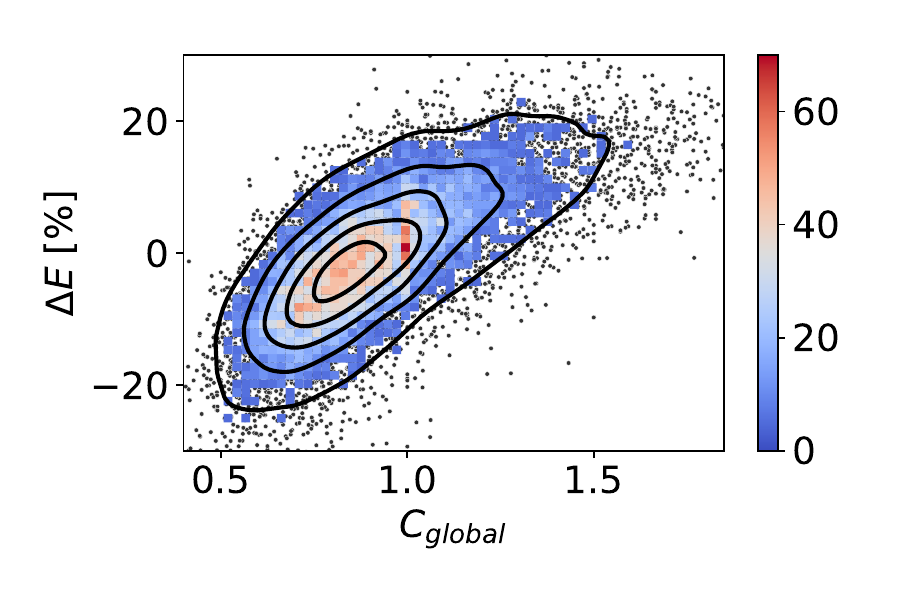}
    \caption{}
    \label{fig:Cglobal_2dCorrelation}
  \end{subfigure}
  \caption{Correlation of \textbf{(a)} the fraction of late energy deposits and \textbf{(b)}
  $C_\text{global}$ with the ratio of the reconstructed to the true beam energy ($\Delta E$).
  The amount of entries in the highest density bins holding \SI{90}{\percent} of the data
  are indicated by the color scale, while the remaining lower density bins are
  shown as individual data points (gray). Additionally, contour lines generated by a
  kernel density estimation are drawn in black.}
  \label{fig:2dCorrelation}
\end{figure}

\autoref{fig:Tedep_2dCorrelation} shows the 2D-distribution of the fraction of late
energy depositions in an event and the relative difference of the reconstructed energy to
the beam energy ($\Delta E$) for \SI{40}{\GeV} pions. It can be seen that for low values of the fraction of late energy deposits ($E_\text{dep} (T_\text{hit} > \SI{3}{\ns})$)
on average the reconstructed energy is too high, while for high fractions of late
energy depositions the reconstructed energy tends to be too low. This observation is in line with the interpretation that a large amount of late energy depositions indicates a high hadronic fraction and thus a lower calorimeter response. For comparison in \autoref{fig:Cglobal_2dCorrelation} $\Delta E$ is shown dependent on $C_\text{global}$, an observable that describes the hit energy distribution of an event and which was developed  to correct the reconstructed energy within the AHCAL \cite{Adloff:2012gv} in the global software compensation scheme. This variable is defined as the ratio between
the fraction of hits in the calorimeter with energy depositions larger than \SI{5}{\text{MIP}} ($C_\text{thr}$) and the fraction of hits with energy depositions larger than the average hit energy in this shower ($C_\text{av}$):
\begin{equation}
  C_\text{global} = \frac{C_\text{thr}}{C_\text{av}}.
\end{equation}
This variable exhibits a positive correlation with $\Delta E$, which was exploited in  \cite{Adloff:2012gv} to improve the energy resolution of the calorimeter.

\begin{figure}[ht]
  \center
    \includegraphics[width = 0.495\textwidth]{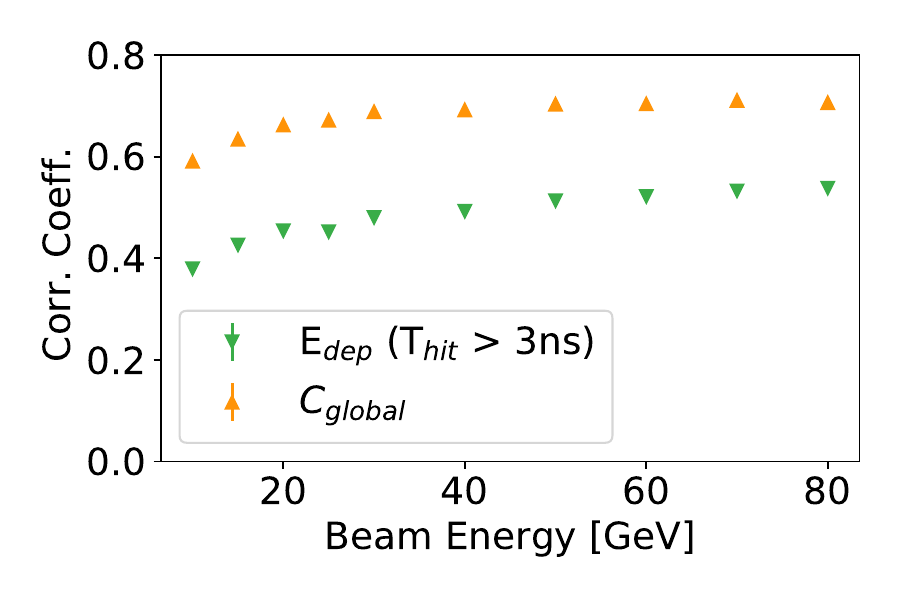}
    \caption{Pearson correlation coefficients of the reconstructed energy
    ($E_\text{std}$, see \autoref{sec:methods}) with the fraction of late energy deposits
    (green) and $C_\text{global}$ (orange).}
    \label{fig:correlationCoeff}
\end{figure}

In \autoref{fig:correlationCoeff} the absolute value of the correlation coefficients between
the fraction of late energy hits and $C_\text{global}$ with $\Delta E$ are shown over
the full energy range. Significant correlations of both observables with the reconstructed
energy are visible. For $C_\text{global}$, however, the correlation is consistently higher
than for the time-based observable. For higher energies the correlation coefficients
are slightly higher, because of the lower statistical uncertainty resulting from the larger number of hits in these events. The large correlation
coefficients observed indicate that both observables can be used to enhance the energy resolution of the AHCAL.

\begin{figure}[ht]
\captionsetup[subfigure]{aboveskip=1pt}
  \center
  \begin{subfigure}{0.495\textwidth}
    \includegraphics[width = 1\textwidth]{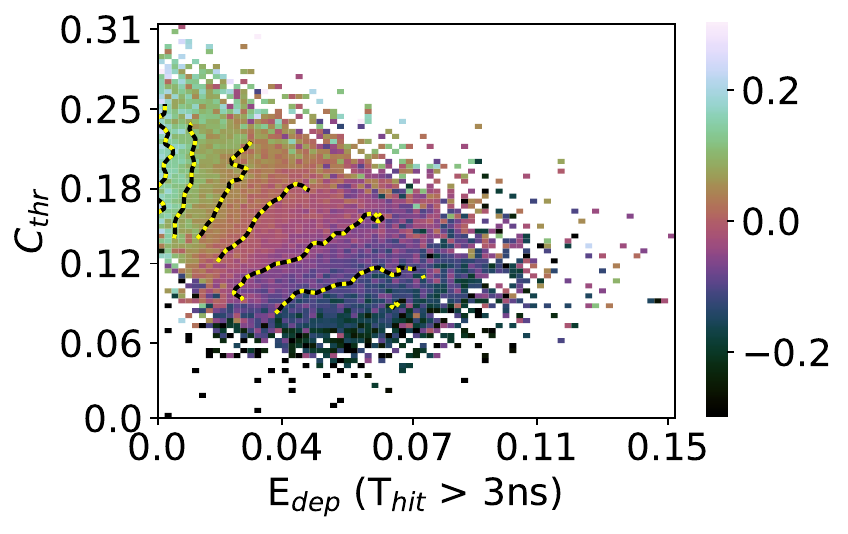}
    \caption{}
    \label{fig:3dCorrelations_Cthr}
  \end{subfigure}
  \hfill
  \begin{subfigure}{0.495\textwidth}
    \includegraphics[width = 1\textwidth]{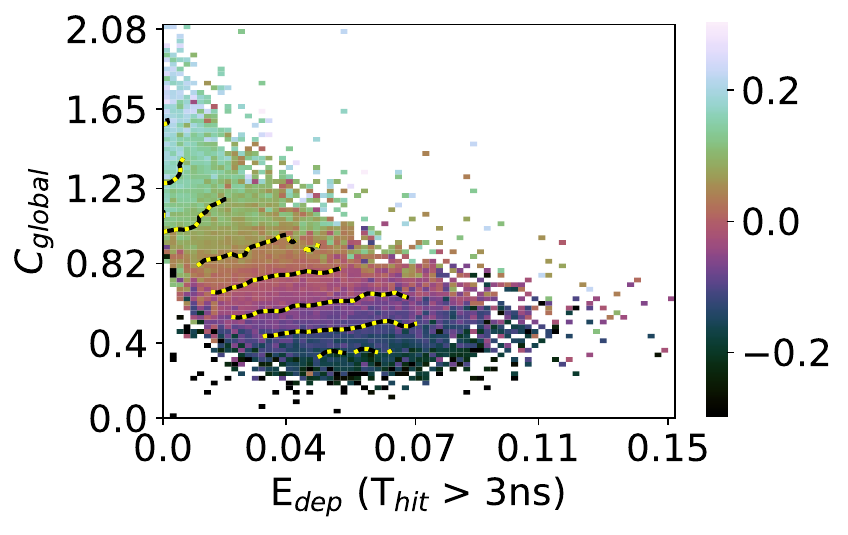}
    \caption{}
    \label{fig:3dCorrelations_Cglobal}
  \end{subfigure}
  %\vspace{-0.4\baselineskip}
  \caption{Mean relative deviation of the reconstructed and true energy for \SIrange{30}{50}{\GeV} pions in each bin shown by the color scale. The data is binned with respect to the fraction of late  energy deposits ($E_\text{dep} (T_\text{hit} > \SI{3}{\ns})$) and \textbf{(a)}  $C_\text{thr}$ and \textbf{(b)} $C_\text{global}$. The yellow dotted line indicates the smoothed contour lines of the color space.
  }
  \label{fig:3dCorrelations}
\end{figure}

In order to study the interplay of the correlation of the late energy depositions
and $C_\text{global}$ with $\Delta E$, we decompose the $C_\text{global}$ variable
and look at the correlations with $\Delta E$ separately. The numerator in the
definition of $C_\text{global}$ is $C_\text{thr}$ describing the fraction of the hits
in a shower that are of high energy density, thus being sensitive to the electromagnetic
part of the shower. The fraction of late energy deposits is sensitive to the amount
of neutron interaction in the shower. In \autoref{fig:3dCorrelations_Cthr} the data
of \SIrange{30}{50}{\GeV} pions is binned with respect to both variables, $C_\text{thr}$
and $E_\text{dep} (T_\text{hit} > \SI{3}{\ns})$. Additionally the deviation of the
reconstructed to the true energy ($\Delta E$) is indicated as the color scale.
Areas of lighter color indicate a too high energy reconstruction, while areas of
darker color indicate a too low reconstructed energy. The contour lines of the
color space are indicated as a yellow dotted line. It can be seen that the contour
lines are not horizontal and thus both variables are necessary to explain the variance
in $\Delta E$. Both information of the size of the electromagnetic part as well as
of the neutron activity are necessary for an optimal explanation of $\Delta E$.

In \autoref{fig:3dCorrelations_Cglobal} the data is binned with respect to $C_\text{global}$
on the $y$-axis. In this case, the contour lines are nearly parallel to the $x$-axis
indicating that $C_\text{global}$ incorporates information about the electromagnetic part of the shower as well as the neutron activity and thus the timing information only
adds marginal value for an improved reconstruction of the energy with respect to the information encoded in $C_\text{global}$.

\section{Methods}
\label{sec:methods}

The standard energy reconstruction method
of a hadronic shower $i$ is the sum of all hit energies $e_j$:

\begin{equation}
    E_{\text{std},i} = \sum_{j \, \in \, \text{hits}} e_{j,i}.
    \label{eq:Estd}
\end{equation}

The software compensation methods improve on this simple reconstruction by
learning weights to down- or upweight certain parts of the shower.
Two different methods are discussed for the time assisted energy reconstruction
in the following: A global and a local software compensation technique.

\subsection{Global software compensation}
The global software compensation technique corrects the energy sum of the
standard reconstruction $E_\text{std}$ by a global correction factor, which is a
function of an observable $\theta$ of the shower. In previous work the aforementioned
$C_\text{global}$ variable was used as the observable $\theta$ \cite{Adloff:2012gv}.
The correction factor is a second order polynomial of $\theta$, such that the
reconstructed energy of event $i$ becomes:

\begin{equation}
    E_{\text{reco},i}^\text{global} = E_{\text{std},i} \cdot (a+b\,\theta_i+c\,\theta_i^2).
    \label{eq:globalSC_1}
\end{equation}
The parameters $a$, $b$, and $c$ are determined by an optimization procedure
minimizing the following loss function:

\begin{equation}
    L = \sum_{i \, \in \, \text{events}} \frac{\left( E_{\text{reco},i}^\text{global} -
        E_{\text{beam},i}\right)^2}{E_{\text{beam},i}},
\end{equation}
where the sum runs over all events in the sample and $E_\text{beam}$ is the simulated
true beam energy. In order to incorporate the hit time measurements into this approach,
it can be extended by using a second observable $\phi$:

\begin{equation}
    E_{\text{reco},i}^\text{global} = E_{\text{std},i} \cdot (a+b\,\theta_i+c\,\phi_i+d\,\theta_i\phi_i),
    \label{eq:globalSC_2}
\end{equation}

where, in this case, $\phi_i$ is the fraction of late energy depositions in a shower.
In order to limit the overall number of parameters the quadratic terms are omitted
in this approach, but a correlation term $\theta_i\phi_i$ is added.
The additional improvement from the quadratic terms on the energy reconstruction is much smaller compared to that  achieved with the already considered terms.

As the parameters $a$ to $d$ are expected to change with the beam energy they are
determined for every training beam energy separately. The resulting parameters
are then fitted with a quadratic function such that:

\begin{equation}
  a(E) = a_0 + a_1 \, E + a_2 \, E ^2,
\end{equation}

and for parameters $b$ to $d$ accordingly. While in the training the true beam energy
is used to determine parameters $a_0$ to $a_2$, for the test data set the standard
energy reconstruction $E_{\text{std},i}$ is used to determine the parameters $a$ to
$d$ for each shower separately. In total the global software compensation method
has nine free parameters in the case of using one observable (\autoref{eq:globalSC_1})
and twelve free parameters in the case of using two observables (\autoref{eq:globalSC_2}).

\subsection{Local software compensation}
In the local software compensation approach, weights are applied separately to each
hit in the shower following the procedure described in \cite{Adloff:2012gv}. For each shower, the hits are binned with respect to their energy $e$ and different weights $w$ are applied for
each bin:

\begin{equation}
        E_\text{reco}^\text{local} = \sum_{j \, \in \, \text{hits}} e_j \cdot w(e_j, E_\text{std}).
\end{equation}
The optimal weights depend on the shower energy. Following the approach in \cite{CALICE:2018ibt}, the weights in each bin are parameterized by a second order polynomial of the energy:

\begin{equation}
  w(e_j, E) = e_j \cdot (c_{k, 0} + c_{k, 1}\,E + c_{k, 2}\,E^2),
\end{equation}
with $c_k$ being the set of parameters for the $k$th energy bin of the corresponding hit energy $e_j$.
During training the true beam energy $E_{\text{beam},i}$ is used in the above equation,
while for testing the standard reconstructed energy $E_{\text{std}}$ is used
to evaluate the weights. The set of weights is determined by minimizing the
following loss function:

\begin{equation}
        L = \sum_{i \, \in \, \text{events}} \frac{\left[\sum_{j \, \in \,
        \text{hits}} e_{j,i} \cdot w(e_{j,i}, E_{\text{beam},i}) - E_{\text{beam},i}\right]^2}{E_{\text{beam},i}}.
\end{equation}

In the presented analysis the hit energy bins shown in \autoref{tab:localSCBins}) are used. 

\begin{table}[h!]
  \caption{Energy range of bins used in the local software compensation approach.}
  \begin{tabular}{l|cccccccc}
    Bin                 & 0 & 1 & 2 & 3 & 4 & 5 & 6 & 7 \\
    \hline
    Range [MIP]  & \numrange[range-phrase = --]{0.5}{1.0}
                        & \numrange[range-phrase = --]{1.0}{1.5}
                        & \numrange[range-phrase = --]{1.5}{2.5}
                        & \numrange[range-phrase = --]{2.5}{5.0}
                        & \numrange[range-phrase = --]{5.0}{15}
                        & \numrange[range-phrase = --]{15}{50}
                        & \numrange[range-phrase = --]{50}{100}
                        & $> 100$ \\
  \end{tabular}
  \label{tab:localSCBins}
\end{table}

In order to incorporate hit time measurements, the total number of bins is doubled. One
set of hit energy bins is used for early hits ($\leq \SI{3}{\ns}$), and the other
one for late hits ($> \SI{3}{\ns}$). The number of free parameters is three per bin, resulting in 24 free parameters for the local software compensation without timing, and 48 with timing included. 

\section{Results}
\begin{figure}[ht]
  \center
  \begin{subfigure}{0.495\textwidth}
    \includegraphics[width = 1\textwidth]{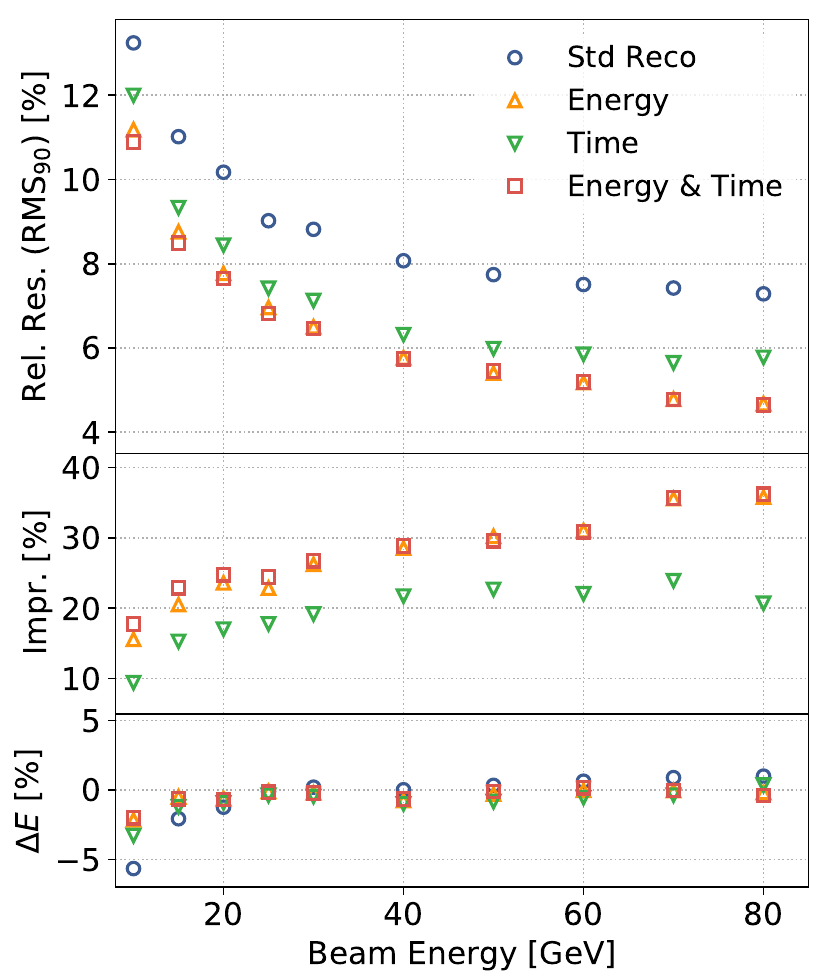}
    \caption{Global software compensation}
    \label{fig:results:globalSC}
  \end{subfigure}
  \hfill
  \begin{subfigure}{0.495\textwidth}
    \includegraphics[width = 1\textwidth]{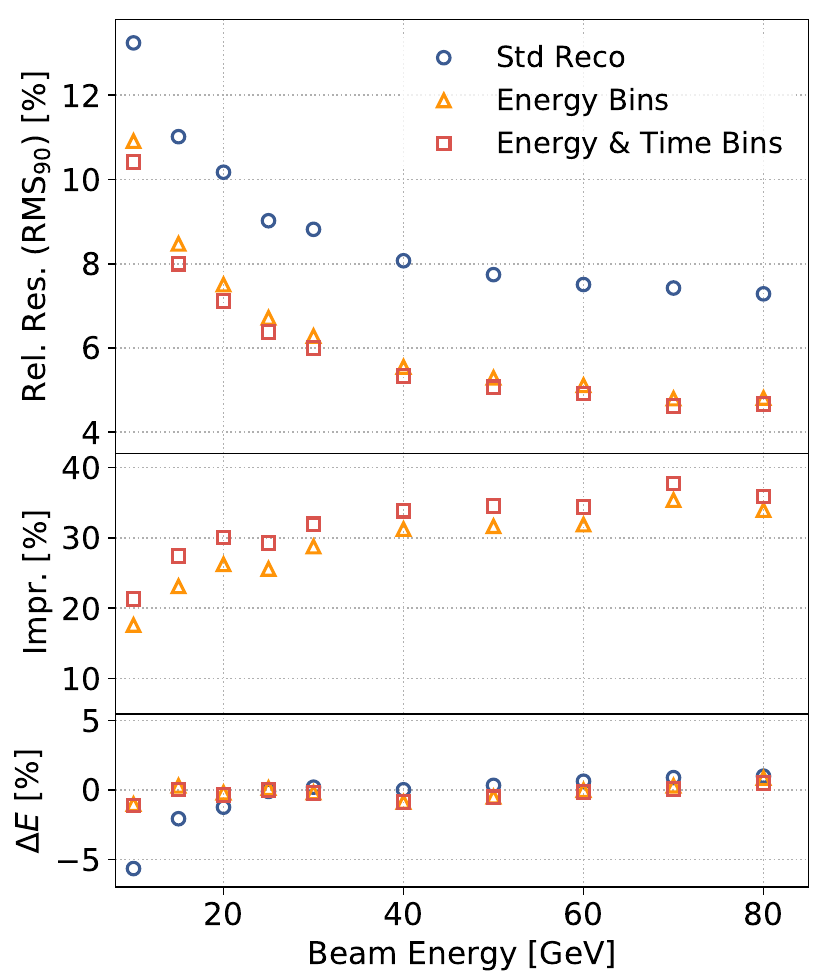}
    \caption{Local software compensation}
    \label{fig:results:localSC}
  \end{subfigure}
  \caption{Results of using \textbf{(a)} global and \textbf{(b)} local software
  compensation using different observables compared to the standard reconstruction.}
  \label{fig:results}
\end{figure}

The results of using the global and the local software compensation approach
are presented in \autoref{fig:results}. The figures consist of three panels with the
beam energy on the $x$-axis. The first panel shows the relative resolution measured
by the $\text{RMS}_{90}$, which is defined as the root mean squared of the smallest
\SI{90}{\percent} interval of the reconstructed energy distributions\footnote{$\text{RMS}_{90}$ is adopted here to provide a robust measure of the width of the distribution while reducing the sensitivity to tails. For a perfectly gaussian distribution, $\text{RMS}_{90}$ is 21\% smaller than the $\sigma$ of the distribution.}. The second panel
shows the improvement in the $\text{RMS}_{90}$ of the software compensation methods with
respect to the standard reconstruction. The third panel indicates the linearity of
the methods shown by $\Delta E$, which is the relative deviation of the median
reconstructed energy compared to the true beam energy. Statistical uncertainties are at or below the size of the markers. This is also true for similar figures below.

In \autoref{fig:results:globalSC} it can be
seen that correcting the reconstructed energy globally by means of the fraction
of late energy deposits, improvements of up to \SI{25}{\percent} over the standard
reconstruction are achieved. This shows that the correlations observed in
\autoref{sec:correlations} can be used to obtain a significant improvement in energy resolution.
However, using the same approach with the $C_\text{global}$ observable leads to larger
improvements of up to \SI{35}{\percent}. The combination of both observables leads
only to minor further improvements in the energy range below \SI{30}{\GeV}. This
is expected because of the observations made in \autoref{sec:correlations}. The observed benefits of the use of time information at lower energies are attributed to larger fluctuations in the shower observables at lower energies, as the number of hits that enter into the event-wise calculation of $C_\text{global}$ is lower. The additional timing information can be exploited for a more accurate estimate on the correction factor of the reconstructed energy. This effect diminishes with increasing energy.

For the local approach to software compensation using only energy bins leads to improvements
of about \SI{35}{\percent} over the standard reconstruction. At the highest energy points this is slightly less than
what is achieved for the global approach, however, for energies of 50 GeV and below
the improvements are larger for the local software compensation.
Adding binning with regards to the hit times improves the energy resolution further by
about \SIrange{3}{4}{\percent} on average, bringing it to or beyond the level achieved with global software compensation over the full energy range. This shows that on a local level the hit time
measurements are capable of improving the energy resolution compared to the standard
local software compensation approach used for the energy reconstruction in the AHCAL. The linearity of both the local and global software compensation approaches are improved compared to the standard reconstruction and comparable between the purely energy based and time-assisted approaches. 

\begin{figure}[ht]
  \center
    \includegraphics[width = 0.495\textwidth]{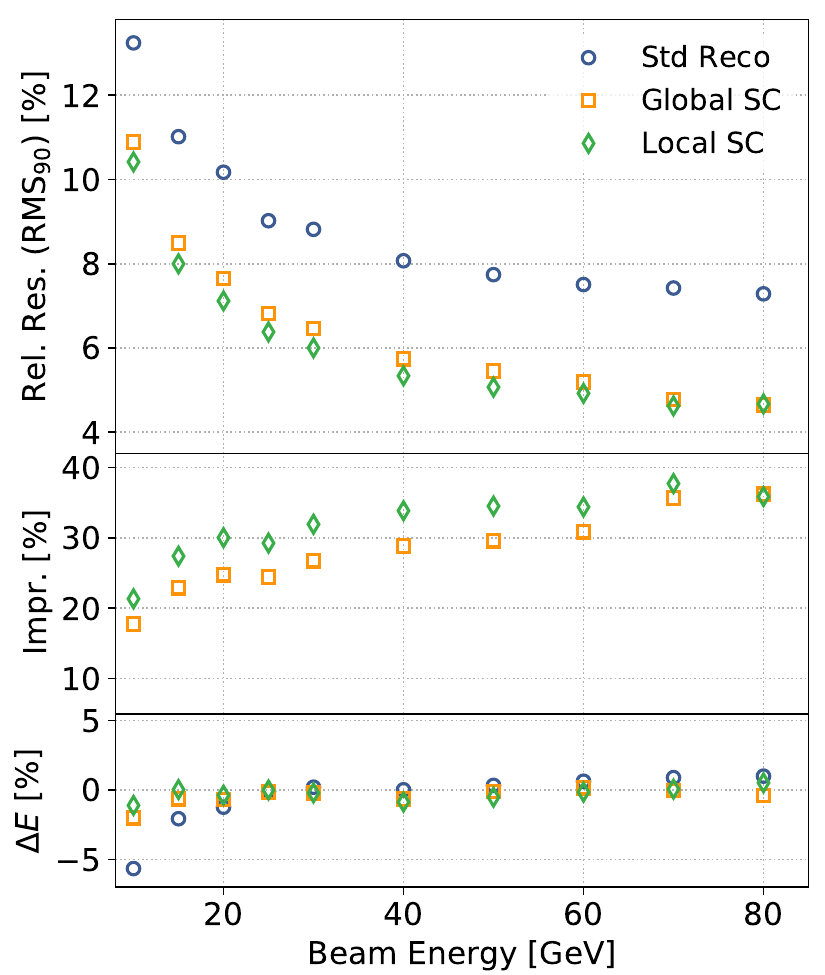}
    \caption{Comparison of global and local software compensation using both energy and time observables.}
    \label{fig:results:globalLocalSC}
\end{figure}

In \autoref{fig:results:globalLocalSC} the global and local version of the time
assisted software compensation methods are compared. Besides the highest energy
of \SI{80}{\GeV} where both methods show the same performance, the local approach gives better results with about a \SI{5}{\percent}
higher improvement over the standard reconstruction for most energies. These results
reflect the observation that the local approach to software compensation is making
better use of the additional time information.

\subsection{Weights for Local Software Compensation}
\begin{figure}[ht]
  \center
  \begin{subfigure}{0.495\textwidth}
    \includegraphics[width = 1\textwidth]{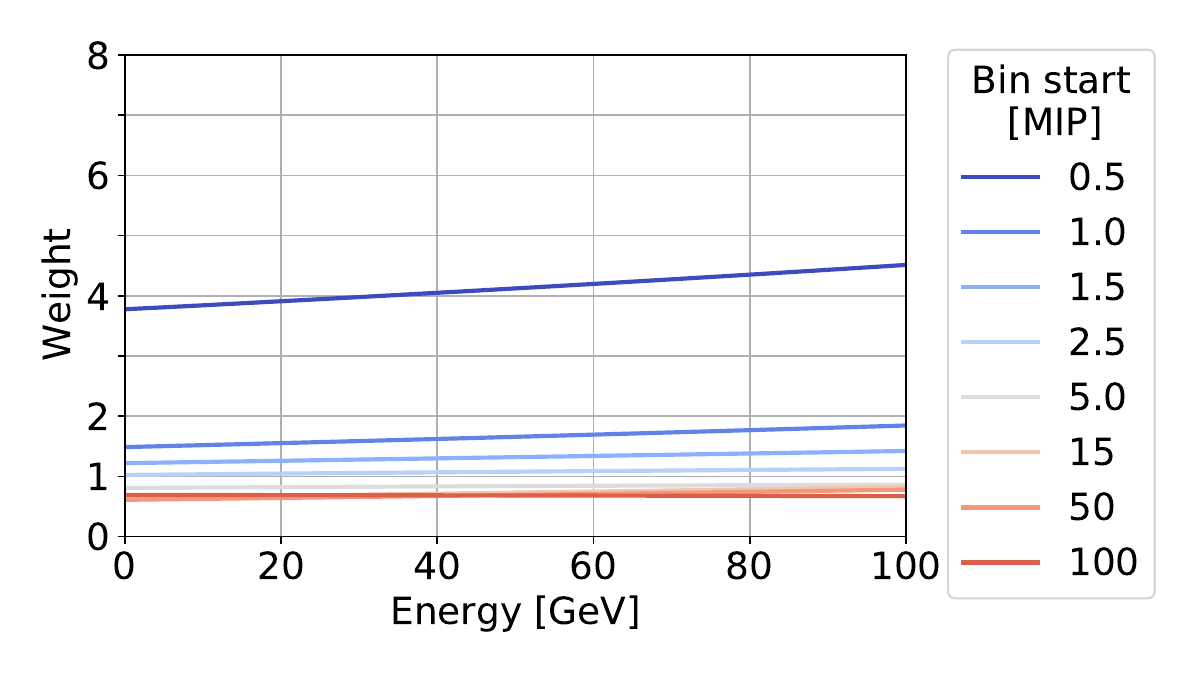}
  \end{subfigure}
  \caption{Weights for the local software compensation approach using only one set of
    energy bins.
    The different colors represent the weights for the different hit energy bins with
    the lower edge of the corresponding bin indicated in the legend.}
  \label{fig:weights_localSC_energy}
\end{figure}

\autoref{fig:weights_localSC_energy} shows the weights used for the local software compensation
approach which uses only one set of hit energy bins. The weights of the first four bins are above one
for the whole energy range, meaning hit energies in these bins get weighted up. The
weights for the other bins are consistently below one and therefore energy depositions
in these bins are weighted down.

\begin{figure}[ht]
  \center
  \begin{subfigure}{0.495\textwidth}
    \includegraphics[width = 1\textwidth]{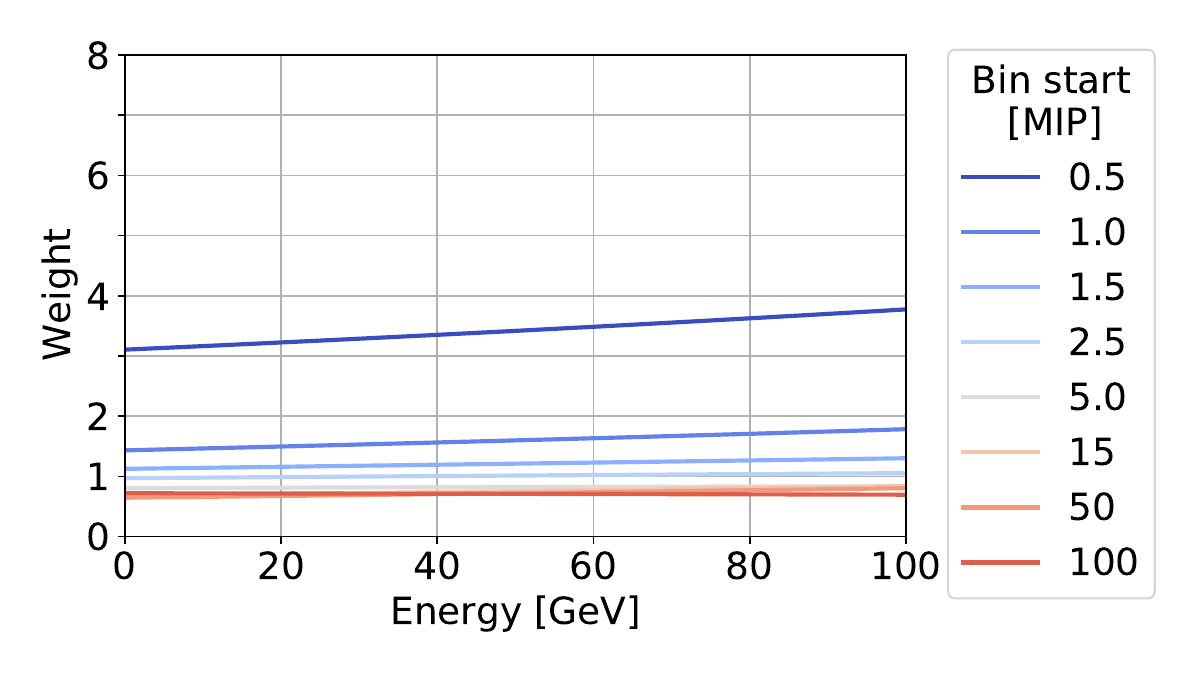}
    \caption{Early hits}
    \label{fig:weights_localSC_energyTime_early}
  \end{subfigure}
  \hfill
  \begin{subfigure}{0.495\textwidth}
    \includegraphics[width = 1\textwidth]{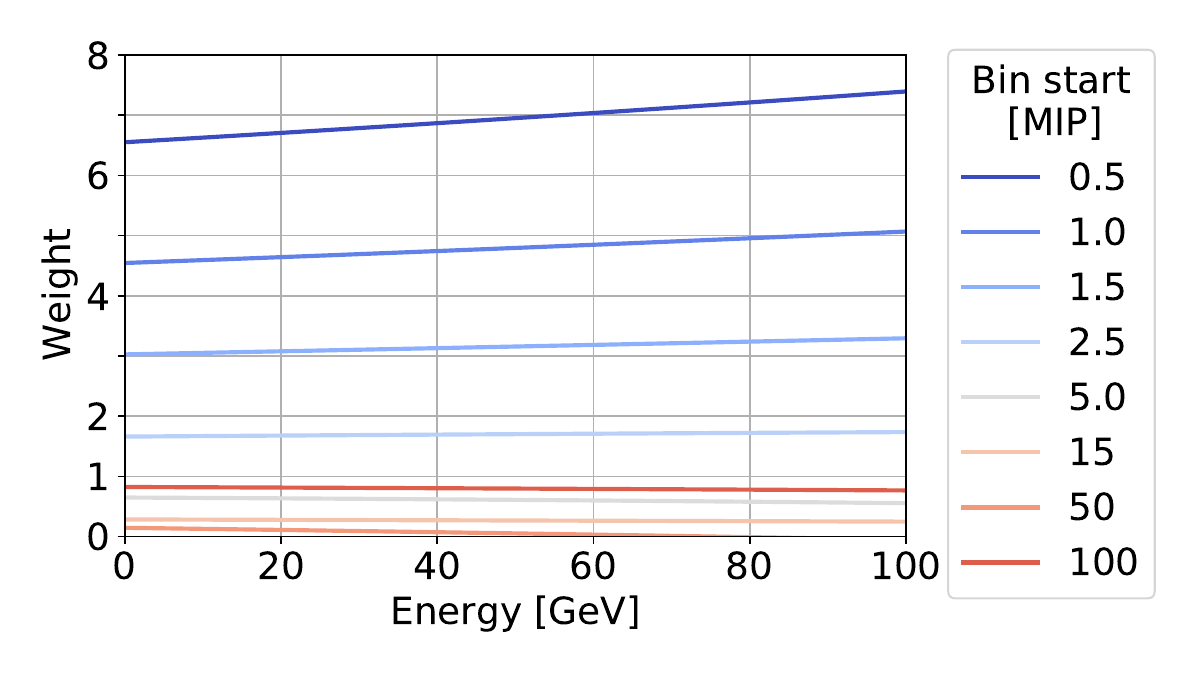}
    \caption{Late hits}
    \label{fig:weights_localSC_energyTime_late}
  \end{subfigure}
  \caption{Weights for the local software compensation approach using energy bins for
    \textbf{(a)} early ($\leq \SI{3}{\ns}$) and \textbf{(b)} late ($> \SI{3}{\ns}$) hits.
    The different colors represents the weights for the different hit energy bins with
    the start of the corresponding bin indicated in the legend.}
  \label{fig:weights_localSC_energyTime}
\end{figure}

The weights for the local software compensation approach using two sets of hit energy bins
is shown in \autoref{fig:weights_localSC_energyTime} for early and late hits. It can
be seen that the low energy hits in the late bins are weighted up much more compared
to the early energy bins. The late energy bins greater than $\SI{5}{\text{MIP}}$ include only very
few data samples, as high-energetic, late energy depositions are rare in the detector.
It can be concluded that the improvement in the resolution of the energy reconstruction
process for the time assisted local software compensation is due to the power of
distinguishing early and late low-energetic energy depositions and the corresponding increased
weighting of the late energy depositions.

\section{Impact of a limited time resolution}

To explore the impact of the limited time resolution of \SI{1}{ns} on the performance of the reconstruction algorithms discussed here, the reconstruction methods are repeated assuming a perfect time resolution. In this case, the time of a calorimeter hit is given by the exact time of the first energy deposition in that particular cell in the simulations, without additional time smearing in the digitization step. All other digitization effects are applied. The parameters of the analysis, in particular the value of \SI{3}{ns} used to separate early and late shower components, are the same. 

\begin{figure}[ht]
  \center
  \begin{subfigure}{0.495\textwidth}
    \includegraphics[width = 1\textwidth]{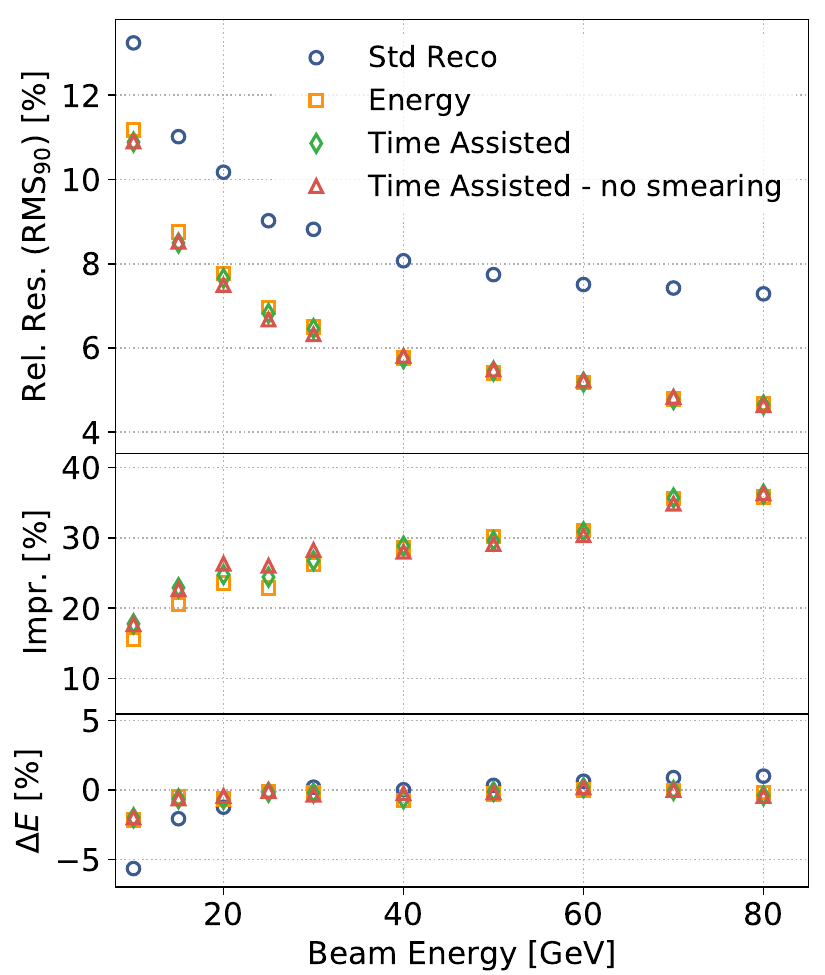}
    \caption{Global software compensation}
    \label{fig:PerfectTiming_global}
  \end{subfigure}
  \hfill
  \begin{subfigure}{0.495\textwidth}
    \includegraphics[width = 1\textwidth]{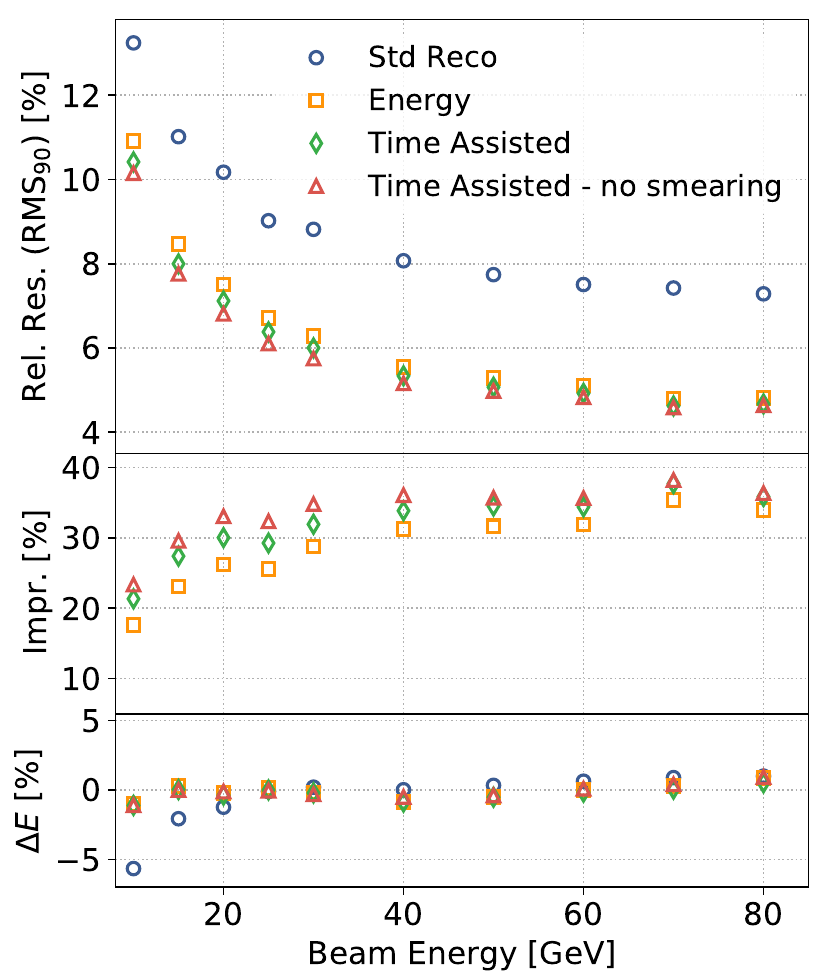}
    \caption{Local software compensation}
    \label{fig:PerfectTiming_local}
  \end{subfigure}
  \caption{Results of \textbf{(a)} global and \textbf{(b)} local software
  compensation using different observables, and assuming a currently realistic time resolution of \SI{1}{\ns} and an ideal time resolution without time smearing, compared to the standard reconstruction.}
  \label{fig:PerfectTiming}
\end{figure}

The results are shown in \autoref{fig:PerfectTiming}. For global software compensation (\autoref{fig:PerfectTiming_global}) no improvement of the energy resolution can be observed with an ideal time resolution compared to a limited time resolution of \SI{1}{\ns}. The energy resolution is dominated by the contribution of $C_\text{global}$. The more accurate time information does not add to a better performance. For local software compensation a clear improvement in energy resolution can be seen (\autoref{fig:PerfectTiming_local}). The improvement in energy resolution compared to the standard reconstruction increases by \SIrange{2}{3}{} percent points for energies below \SI{50}{\GeV}. The better separation between early and late hits has a direct impact on the performance of the algorithm.

\section{Conclusions}
Two methods of time assisted energy reconstruction within the CALICE \ac{AHCAL}
were presented and studied on simulated data. Both methods rely on the time structure
of hadronic showers, where late hits are predominantly caused by late neutrons. Strong
correlations of the fraction of late energy deposits with the reconstructed energy were
observed. In a global software compensation approach these correlations have been exploited
for an improved energy resolution compared to the standard reconstruction. However,
the improvement is less than what can be achieved using an observable based on the
hit energy distribution. The combination of both time and energy density observables
does not lead to significant improvements over the purely energy-based methods.
For a local software compensation approach the number of bins were doubled and weights
have been derived separately for early and late energy depositions. With this approach,
improvements in the energy resolution compared to the standard local software
compensation approach using only energy density have been observed. Over most of the energy range, a better energy resolution than for the global software compensation has been achieved. A perfect time resolution enables additional improvements compared to a \SI{1}{ns} time resolution. 

These results show that time information in highly-granular hadronic calorimeters, while correlated with information provided by energy density, is beneficial for energy reconstruction, in particular when local information on the cell level is used. Considering that the local software compensation technique introduced here uses only two bins in time, it is expected that more complex algorithms based on advanced machine-learning techniques which can use the full time information in a five-dimensional reconstruction (space, energy density, and time) will enable further improvements in energy reconstruction. Studies such as \cite{Akchurin:2021afn} give first impressions of this potential.

\section*{Acknowledgements}
The authors thank the CALICE Collaboration for fruitful discussions and helpful comments. This work would not have been possible without the use of the CALICE software framework.

\bibliographystyle{JHEP}
\bibliography{references}

\providecommand{\href}[2]{#2}\begingroup\raggedright\begin{thebibliography}{10}

\bibitem{Thomson:2009rp}
M.~A. Thomson, \emph{{Particle Flow Calorimetry and the PandoraPFA Algorithm}},
  \href{http://dx.doi.org/10.1016/j.nima.2009.09.009}{\emph{Nucl. Instrum.
  Meth. A} {\bfseries 611} (2009) 25--40},
  [\href{https://arxiv.org/abs/0907.3577}{{\ttfamily arXiv:0907.3577}}].

\bibitem{Brient:2001fow}
J.-C. Brient and H.~Videau, \emph{{The Calorimetry at the future e+ e- linear
  collider}}, {\emph{eConf} {\bfseries C010630} (2001) E3047},
  [\href{https://arxiv.org/abs/hep-ex/0202004}{{\ttfamily
  arXiv:hep-ex/0202004}}].

\bibitem{pfaMorgunov}
V.~L. Morgunov, \emph{{Calorimetry design with energy-flow concept (imaging
  detector for high-energy physics)}},  in \emph{{CALOR 2002, Pasadena,
  California}}, 2002.

\bibitem{CALICE:2010fpb}
{\scshape CALICE} collaboration, C.~Adloff et~al., \emph{{Construction and
  Commissioning of the CALICE Analog Hadron Calorimeter Prototype}},
  \href{http://dx.doi.org/10.1088/1748-0221/5/05/P05004}{\emph{JINST}
  {\bfseries 5} (2010) P05004},
  [\href{https://arxiv.org/abs/1003.2662}{{\ttfamily arXiv:1003.2662}}].

\bibitem{Tran:2017tgr}
H.~L. Tran, K.~Kr\"uger, F.~Sefkow, S.~Green, J.~Marshall, M.~Thomson et~al.,
  \emph{{Software compensation in Particle Flow reconstruction}},
  \href{http://dx.doi.org/10.1140/epjc/s10052-017-5298-3}{\emph{Eur. Phys. J.
  C} {\bfseries 77} (2017) 698},
  [\href{https://arxiv.org/abs/1705.10363}{{\ttfamily arXiv:1705.10363}}].

\bibitem{Abramowicz:1980iv}
H.~Abramowicz et~al., \emph{{The Response and Resolution of an Iron
  Scintillator Calorimeter for Hadronic and Electromagnetic Showers Between
  10-{GeV} and 140-{GeV}}},
  \href{http://dx.doi.org/10.1016/0029-554X(81)90083-5}{\emph{Nucl. Instrum.
  Meth.} {\bfseries 180} (1981) 429}.

\bibitem{H1CalorimeterGroup:1993uzc}
{\scshape H1 Calorimeter Group} collaboration, B.~Andrieu et~al.,
  \emph{{Results from pion calibration runs for the H1 liquid argon calorimeter
  and comparisons with simulations}},
  \href{http://dx.doi.org/10.1016/0168-9002(93)91258-O}{\emph{Nucl. Instrum.
  Meth. A} {\bfseries 336} (1993) 499--509}.

\bibitem{ATLASLiquidArgonEMECHEC:2004hkk}
{\scshape ATLAS Liquid Argon EMEC/HEC} collaboration, C.~Cojocaru et~al.,
  \emph{{Hadronic calibration of the ATLAS liquid argon end-cap calorimeter in
  the pseudorapidity region 1.6 \ensuremath{<} |$\eta$| \ensuremath{<} 1.8 in
  beam tests}}, \href{http://dx.doi.org/10.1016/j.nima.2004.05.133}{\emph{Nucl.
  Instrum. Meth. A} {\bfseries 531} (2004) 481--514},
  [\href{https://arxiv.org/abs/physics/0407009}{{\ttfamily
  arXiv:physics/0407009}}].

\bibitem{Adloff:2012gv}
{\scshape CALICE} collaboration, C.~Adloff et~al., \emph{{Hadronic energy
  resolution of a highly granular scintillator-steel hadron calorimeter using
  software compensation techniques}},
  \href{http://dx.doi.org/10.1088/1748-0221/7/09/P09017}{\emph{JINST}
  {\bfseries 7} (2012) P09017},
  [\href{https://arxiv.org/abs/1207.4210}{{\ttfamily arXiv:1207.4210}}].

\bibitem{Sefkow:2018rhp}
{\scshape CALICE} collaboration, F.~Sefkow and F.~Simon, \emph{{A highly
  granular SiPM-on-tile calorimeter prototype}},
  \href{http://dx.doi.org/10.1088/1742-6596/1162/1/012012}{\emph{J. Phys. Conf.
  Ser.} {\bfseries 1162} (2019) 012012},
  [\href{https://arxiv.org/abs/1808.09281}{{\ttfamily arXiv:1808.09281}}].

\bibitem{Emberger:2021lsz}
L.~Emberger, \emph{{Analysis of Testbeam Data Recorded with the Large CALICE
  AHCAL Technological Prototype}},  in \emph{{International Workshop on Future
  Linear Colliders}}, 5, 2021,
  \href{https://arxiv.org/abs/2105.08497}{{\ttfamily arXiv:2105.08497}}.

\bibitem{WA78:1985num}
{\scshape WA78} collaboration, M.~De~Vincenzi et~al., \emph{{Experimental Study
  of Uranium Scintillator and Iron Scintillator Calorimetry in the Energy Range
  135-{GeV} to 350-{GeV}}},
  \href{http://dx.doi.org/10.1016/0168-9002(86)90968-X}{\emph{Nucl. Instrum.
  Meth. A} {\bfseries 243} (1986) 348}.

\bibitem{Caldwell:1992te}
A.~Caldwell, L.~Hervas, J.~A. Parsons, F.~Sciulli, W.~Sippach and L.~Wai,
  \emph{{Measurement of the time development of particle showers in a uranium
  scintillator calorimeter}},
  \href{http://dx.doi.org/10.1016/0168-9002(93)90568-3}{\emph{Nucl. Instrum.
  Meth. A} {\bfseries 330} (1993) 389--404}.

\bibitem{Acosta:1990bq}
D.~Acosta et~al., \emph{{Electron - pion discrimination with a scintillating
  fiber calorimeter}},
  \href{http://dx.doi.org/10.1016/0168-9002(91)90489-D}{\emph{Nucl. Instrum.
  Meth. A} {\bfseries 302} (1991) 36--46}.

\bibitem{Akchurin:2007uf}
N.~Akchurin et~al., \emph{{Measurement of the Contribution of Neutrons to
  Hadron Calorimeter Signals}},
  \href{http://dx.doi.org/10.1016/j.nima.2007.08.049}{\emph{Nucl. Instrum.
  Meth. A} {\bfseries 581} (2007) 643--650},
  [\href{https://arxiv.org/abs/0707.4019}{{\ttfamily arXiv:0707.4019}}].

\bibitem{Frank:2014zya}
M.~Frank, F.~Gaede, C.~Grefe and P.~Mato, \emph{{DD4hep: A Detector Description
  Toolkit for High Energy Physics Experiments}},
  \href{http://dx.doi.org/10.1088/1742-6596/513/2/022010}{\emph{J. Phys. Conf.
  Ser.} {\bfseries 513} (2014) 022010}.

\bibitem{Agostinelli:2002hh}
{\scshape GEANT4} collaboration, S.~Agostinelli et~al., \emph{{GEANT4--a
  simulation toolkit}},
  \href{http://dx.doi.org/10.1016/S0168-9002(03)01368-8}{\emph{Nucl. Instrum.
  Meth. A} {\bfseries 506} (2003) 250--303}.

\bibitem{Hirschberg:1992xd}
M.~Hirschberg, R.~Beckmann, U.~Brandenburg, H.~Brueckmann and K.~Wick,
  \emph{{Precise measurement of Birks kB parameter in plastic scintillators}},
  \href{http://dx.doi.org/10.1109/23.159657}{\emph{IEEE Trans. Nucl. Sci.}
  {\bfseries 39} (1992) 511--514}.

\bibitem{Bilki:2014bga}
{\scshape CALICE} collaboration, B.~Bilki et~al., \emph{{Pion and proton
  showers in the CALICE scintillator-steel analogue hadron calorimeter}},
  \href{http://dx.doi.org/10.1088/1748-0221/10/04/P04014}{\emph{JINST}
  {\bfseries 10} (2015) P04014},
  [\href{https://arxiv.org/abs/1412.2653}{{\ttfamily arXiv:1412.2653}}].

\bibitem{Adloff:2014rya}
{\scshape CALICE} collaboration, C.~Adloff et~al., \emph{{The Time Structure of
  Hadronic Showers in highly granular Calorimeters with Tungsten and Steel
  Absorbers}},
  \href{http://dx.doi.org/10.1088/1748-0221/9/07/P07022}{\emph{JINST}
  {\bfseries 9} (2014) P07022},
  [\href{https://arxiv.org/abs/1404.6454}{{\ttfamily arXiv:1404.6454}}].

\bibitem{CALICE:2018ibt}
{\scshape CALICE} collaboration, J.~Repond et~al., \emph{{Hadronic Energy
  Resolution of a Combined High Granularity Scintillator Calorimeter System}},
  \href{http://dx.doi.org/10.1088/1748-0221/13/12/P12022}{\emph{JINST}
  {\bfseries 13} (2018) P12022},
  [\href{https://arxiv.org/abs/1809.03909}{{\ttfamily arXiv:1809.03909}}].

\bibitem{Akchurin:2021afn}
N.~Akchurin, C.~Cowden, J.~Damgov, A.~Hussain and S.~Kunori, \emph{{On the use
  of neural networks for energy reconstruction in high-granularity
  calorimeters}},
  \href{http://dx.doi.org/10.1088/1748-0221/16/12/P12036}{\emph{JINST}
  {\bfseries 16} (2021) P12036},
  [\href{https://arxiv.org/abs/2107.10207}{{\ttfamily arXiv:2107.10207}}].

\end{thebibliography}\endgroup

\end{document}